\documentclass{comnet}%%%%where comnet is the template name
\usepackage{amsmath} %for dealing with mathematics,
\usepackage{amsthm} %for dealing with theorem environments,
\usepackage{subfig} %for getting the subfigures e.g., "Figure 1a and 1b" etc.

\usepackage{graphicx}
\usepackage{dcolumn}% Align table columns on decimal point
\usepackage{bm}% textbf math
\usepackage{color}
\usepackage{txfonts}
\usepackage{comment}
\usepackage{longtable}
\usepackage{url}
\usepackage{changepage}
\usepackage{cite}
\usepackage[figuresleft]{rotating}

\begin{document}

\title{Time-dependent community structure in legislation cosponsorship networks in the Congress of the Republic of Peru
}

\shorttitle{Time-dependent community structure in cosponsorship networks} %%%for recto running head
\shortauthorlist{S.\,H. Lee, J.\,M.\ Magallanes, and M.\,A. Porter} %%% for verso running head

\author{%%%% First author details
\name{Sang Hoon Lee$^*$}
\address{School of Physics, Korea Institute for Advanced Study, Seoul 02455, Korea}
\address{Integrated Energy Center for Fostering Global Creative Researcher (BK 21 plus) and Department of Energy Science, 
Sungkyunkwan University, Suwon 16419, Korea}
\address{Oxford Centre for Industrial and Applied Mathematics (OCIAM), Mathematical Institute, University of Oxford, 
Oxford, OX2 6GG, United Kingdom\email{$^*$Corresponding author: lshlj82@kias.re.kr}}
%%%%%%% Second author details
\name{Jos{\'e} Manuel Magallanes}
\address{eScience Institute and Evans School of Public Policy and Governance, University of Washington, WRF Data Science Studio, Physics/Astronomy Tower, 6th Floor, 3910 15th Ave NE, Seattle, WA 98195-1570, USA}
\address{Departamento de Ciencias Sociales and Escuela de Gobierno y Pol{\'i}ticas P{\'u}blicas, Pontificia Universidad Cat{\'o}lica del Per{\'u}, Av.~Universitaria 1801, San Miguel, Lima 32, Peru}
\address{Center for Social Complexity, Krasnow Institute of Advanced Study, George Mason University, 4400 University Dr.,  Fairfax, VA 22030, USA}
%%%%%%%
\and
%%%%%%% Third author details
\name{Mason A. Porter}
\address{Oxford Centre for Industrial and Applied Mathematics (OCIAM), Mathematical Institute, University of Oxford, 
Oxford, OX2 6GG, United Kingdom}
\address{CABDyN Complexity Centre, University of Oxford, Oxford OX1 1HP, United Kingdom}}

\maketitle

%%%%%

\begin{abstract}
{We study community structure in time-dependent legislation cosponsorship networks in the Peruvian Congress, and we compare them briefly to legislation cosponsorship networks in the United States Senate.  
To study these legislatures, we employ a multilayer representation of temporal networks in which legislators in each layer are connected to each other with a weight that is based on how many bills they cosponsor. We then use multilayer modularity maximization to detect communities in these networks. From our computations, we are able to capture power shifts in the Peruvian Congress during 2006--2011.  For example, we observe the emergence of ``opportunists'', who switch from one community to another, as well as cohesive legislative communities whose initial component legislators never change communities. Interestingly, many of the opportunists belong to the group that won the majority in Congress.}

%keywords
{political cosponsorship networks, time-dependent community structure, multilayer networks}
%%%% If classification number provided then
%\\
%2000 Math Subject Classification: 34K30, 35K57, 35Q80,  92D25
\end{abstract}

%%%%%%

\section{Introduction} \label{sec:introduction}

Political networks encompass several types of connectivity --- based on social ties, voting similarities, and other features --- and it is important to analyze them to understand political systems \cite{lazer_networks_2011,ward2011,McClurg2014,victor2016}. In political science, the availability of public data (sometimes aided by various digital media~\cite{SHLee2010}) provides a strong and compelling encouragement for quantitative analyses of politics \cite{pr1997}. In addition to legislative bodies, on which we focus in the present study, numerous types of political networks have now been investigated quantitatively. These include judiciary systems such as the United States (U.S.) Supreme Court~\cite{Fowler2007} and the European Court of Justice~\cite{Mirshahvalad2012}; international relations~\cite{Hafner-Burton2009}; political communication~\cite{lazer_coevolution_2010}; lobbying~\cite{carpenter_friends_2004}; and political behavior~\cite{klofstad_disagreeing_2013}.

As with other types of networks, it has thus far been most common to examine political networks terms of standard (i.e., ``monoplex'') networks, which are represented mathematically as ordinary graphs \cite{NetworkReview}. The types of legislative networks that have been studied in this way include ones defined based on committee assignments \cite{Porter2005,Porter2007,Victor2009,Hedlund2009}, legislation cosponsorship \cite{fowler2006a,fowler2006b,Zhang2008,Kirkland2013,Kirkland2014}, party faction~\cite{Koger2010}, and similarity of voting patterns \cite{SHLee2010,waugh2009,Lyons2009,Alvarez2012,Ringe2012,macon2012,KHahn2013,puck2014,TQPeng2014,Maso2014}. Ideas from temporal networks \cite{TemporalNetwork,holme2013,holme2015} and multilayer networks \cite{mikko,bocca} have been incorporated into investigations of some time-dependent \cite{Mucha2010,mucha2010b,moody2013} and multiplex networks \cite{cranmer2014} in data from politics and international relations. Investigations of the dynamical restructuring of political bodies have yielded insights into the aggregate tendencies of party polarization and realignment \cite{waugh2009,Mucha2010,moody2013} and in the study of politically developing (or democratizing) countries~\cite{calvo_nationalization_2012,aleman_comparing_2009,aleman_explaining_2013,guzman_schrader_alisis_2011}, and using approaches from temporal and multilayer networks promises to generate further insights in these applications.

In the present paper, we consider the dynamical restructuring of a legislature in a democratizing country by examining a time-dependent network of legislators in the Congress of the Republic of Peru~\cite{PeruvianCongress}. We construct a multilayer temporal network using bill cosponsorship relationships among politicians. In our multilayer representation of these temporal networks, the edge weights in each layer arise from the similarities in cosponsorship patterns during the time window that is associated with the layer. Because cosponsorship relationships are one of the main types of work allocated to the legislative branch of the government in Peru, studying them allows us to focus on official activities of politicians instead of either speculative or scandalous ones (social ties, bribery, etc.). To examine how cohesive sets of legislators with cosponsorship relationships change over time, we examine time-dependent community structure~\cite{Mucha2010,CommunityReview}. We find a dramatic rearrangement of community structure in the Peruvian Congress, which we contrast to the relatively stable political bipolarity in the U.S. Senate. For the Peruvian Congress, considering a cosponsorship relationship alone does a good job of successfully revealing the underlying political power rearrangement, and it also appears to deliver more subtle information than official party membership. We quantify the rearrangement of political groups by examining changes over time in time-dependent community structure~\cite{mikko,Mucha2010}. We then do similar computations for legislation cosponsorship networks in the U.S. Senate and compare our results to our observations for the Peruvian Congress.

The rest of our paper is organized as follows.  In Section \ref{sec:method}, we present the data set of Peruvian legislators, discuss how we use it to construct time-dependent networks, and indicate our methodology for analyzing these networks. We discuss our results for the Peruvian Congress and U.S. Senate in Section \ref{sec:results}, and we conclude in Section \ref{sec:conclusion}.

%%%%%%

\section{Data set and methods}
\label{sec:method}

\subsection{Peruvian cosponsorship network data}
\label{sec:data}

\begin{figure*}\centering
\includegraphics[width=0.8\textwidth]{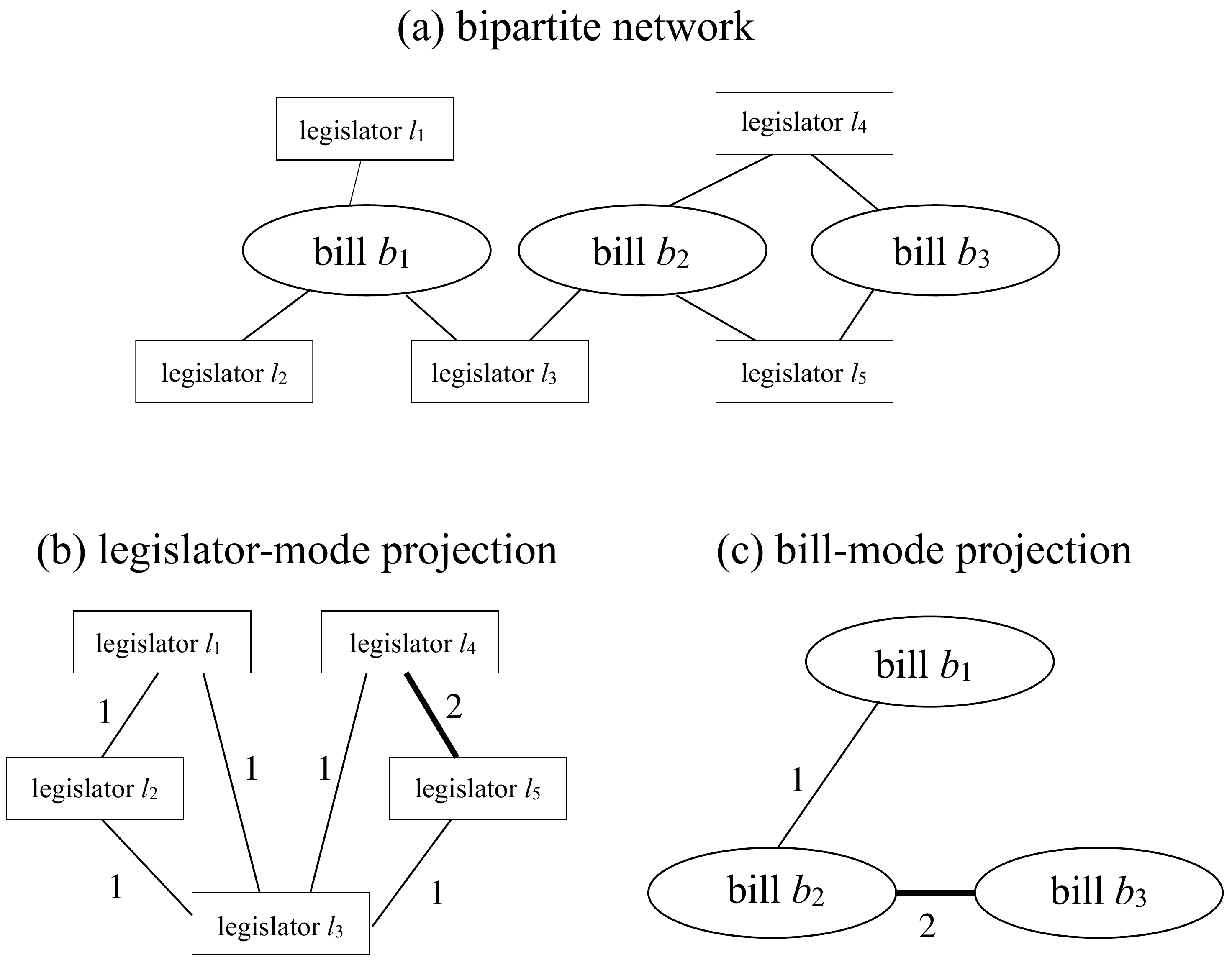}
\caption{Schematic to illustrate a projection from a bipartite network to a weighted network.
The edges are undirected, because we treat all of the main sponsors and their cosponsors equally. (That is, we ignore the
directed nature of such relationships.)
}
\label{bipartite_procedure}
\end{figure*}

\begin{table*}
\caption{Basic statistics for the bipartite cosponsorship networks (BCNs),
bill-mode projection networks (BPNs), and legislator-mode projection networks (LPNs). 
[TW is the time window, \#B is the number of bills, \#L is the number of legislators, \#E is the number of edges, and $\rho$ is the edge density.]
} 
%\begin{ruledtabular}
\centering
%\small
\begin{tabular}{lrrrrr}
\hline
TW & \#B & \#L & \#E\,; $\rho$ & \#E\,; $\rho$ & \#E\,; $\rho$\\
 & & & BCN & BPN & LPN \\
\hline
total & $3\,522$ & $130$ & $27\,414$; $0.0021$ & $1\,569\,063$; $0.253$ & $7\,107$; $0.848$ \\
$0607$\_I & $678$ & $120$ & $5\,251$; $0.0645$ & $56\,411$; $0.246$ & $3\,694$; $0.517$ \\
$0607$\_II & $355$ & $119$ & $2\,824$; $0.0668$ & $18\,447$; $0.294$ & $3\,723$; $0.530$ \\
$0708$\_I & $469$ & $117$ & $3\,656$; $0.0666$ & $25\,891$; $0.236$ & $2\,880$; $0.424$ \\
$0708$\_II & $341$ & $119$ & $2\,629$; $0.0648$ & $13\,418$; $0.231$ & $2\,195$; $0.313$ \\
$0809$\_I & $331$ & $122$ & $2\,550$; $0.0631$ & $12\,163$; $0.223$ & $2\,227$; $0.302$ \\
$0809$\_II & $243$ & $119$ & $2\,032$; $0.0703$ & $8\,244$; $0.280$ & $3\,580$; $0.510$ \\
$0910$\_I & $323$ & $116$ & $2\,394$; $0.0639$ & $11\,644$; $0.224$ & $2\,250$; $0.337$ \\
$0910$\_II & $242$ & $118$ & $1\,942$; $0.0680$ & $7\,655$; $0.263$ & $2\,498$; $0.362$ \\
$1011$\_I & $379$ & $119$ & $2\,875$; $0.0637$ & $18\,320$; $0.256$ & $2\,179$; $0.310$ \\
$1011$\_II & $161$ & $116$ & $1\,261$; $0.0675$ & $3\,612$; $0.280$ & $2\,039$; $0.306$ \\
\hline
\end{tabular}
%\end{ruledtabular}
\label{networks_info_I}
\end{table*}

\begin{figure*}\centering
\begin{tabular}{ll}
(a) & (b) \\
\includegraphics[width=0.5\textwidth]{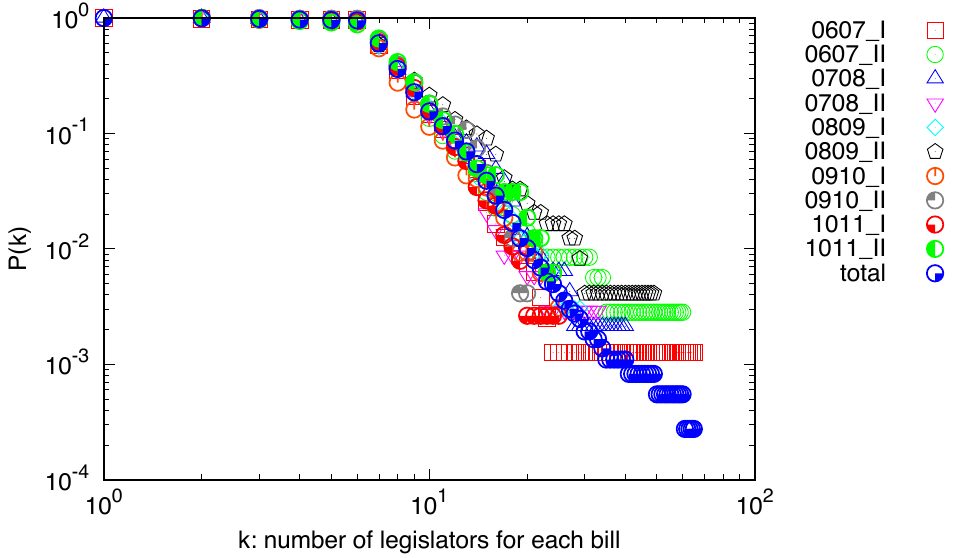} & \includegraphics[width=0.5\textwidth]{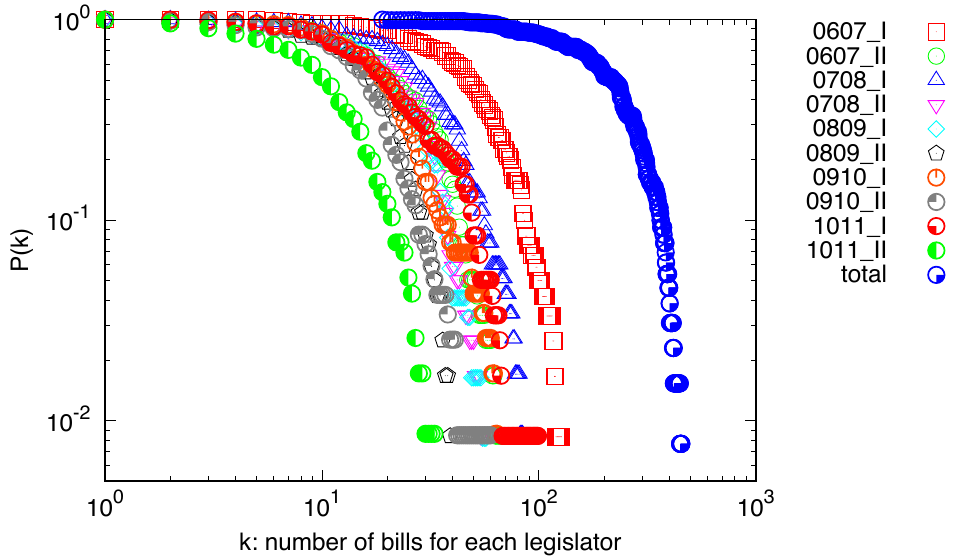} \\
\end{tabular}
\caption{Cumulative degree distributions $P(k) = \sum_{k' \ge k} p(k')$ of (a) bills (i.e., indicating the number of cosponsoring legislators) and (b) legislators (i.e., indicating the number of cosponsored bills) in the bipartite cosponsorship network (BCN) for each half-year. 
}
\label{bipartite_degree_distribution}
\end{figure*}

We manually curated the legislation cosponsorship data for the Peruvian Congress during the years 2006--2011 directly from \cite{PeruvianCongress}.  During 2006--2011, the Peruvian Congress experienced a significant amount of restructuring---including the splitting of the majority {party (Uni\'on por el Per\'u; UPP), which was representing the ``opposition'' to the government}\footnote{In Peru, it is customary that the losing party in the presidential election considers itself to be the ``opposition'' in Congress as a declaration that it is watching the government to make sure that it remains accountable and is not abusing its constitutional powers.}, and the reorganization of legislators from the minority 
{parties}
into different legislative groups\footnote{It is important to distinguish the terms ``party'' and ``group'' in Peruvian politics. A party competes in an election, whereas a group is created in Congress. Naturally, there is a very strong correspondence between parties and groups.}. 
{There were 7 parties/groups at the beginning of the 2006--2011 Congress, 14 groups were formed during that period (though not all were present at one time),}
and there were 9 political groups and 3 legislators without a group affiliation at the end of the Congress in 2011~\cite{GruposParlamentarios}.

A legislative year starts on 27 July every year, so we use ``I'' to the denote bills between that date and the last day of December and ``II'' to denote bills between January and before the beginning of the next legislative year\footnote{This nomenclature helps facilitate understanding of the temporal organization of the bill proposals, but it need not correspond to official legislative sessions, as the official beginnings and endings of sessions can be reduced or extended if the Congress decides to do so. Proposals can be presented during extraordinary sessions, but the webpage of the Congress (whence the data was scraped) does not indicate whether any proposals were presented during an extraordinary session in 2006--2011.}.
Therefore, $0607$\_I occurs in the second half of the year 2006, $0607$\_II occurs during the first half of the year 2007, and so on. A cosponsorship relation occurs between a legislator and a bill, so each half-year constitutes a bipartite (i.e., two-mode) network. As we illustrate in Fig.~\ref{bipartite_procedure}(a), if bill $b_1$ is cosponsored by legislators $l_1$, $l_2$, and $l_3$, then bill $b_1$ is adjacent to $l_1$, $l_2$, and $l_3$ via an undirected edge. For simplicity, we do not distinguish between the main sponsor of a bill and his/her cosponsors, so everyone who participates in sponsoring a bill is adjacent to the bill (i.e., we ignore the directed nature of these edges, as in Ref.~\cite{Zhang2008}). Of course, two bills can share legislators [such as legislator $l_3$ for bills $b_1$ and $b_2$ in Fig.~\ref{bipartite_procedure}(a)], and two legislators can cosponsor multiple bills (such as legislators $l_4$ and $l_5$, who cosponsor bills $b_2$ and $b_3$). Such overlapping relationships are important for ``projections'' of a bipartite cosponsorship network (BCN) [see Figs.~\ref{bipartite_procedure}(b) and \ref{bipartite_procedure}(c)]. Because we are interested in the relationships among the politicians, we use these networks to construct unipartite (i.e., one-mode) networks among legislators [see Fig.~\ref{bipartite_procedure}(b)]. The projection throws away some information~\cite{Porter2005,Porter2007,SHLee2011,everett2013}, but we nevertheless consider such legislator-mode projections as our primary type of network among legislators.

In Fig.~\ref{bipartite_procedure}, we illustrate our procedure to ``project'' from a BCN to weighted, unipartite bill and legislator networks. In the former, a weighted edge indicates the number of cosponsoring legislators that two bills have in common; in the latter, a weighted edge indicates the number of bills that a pair of legislators both cosponsored. We also remove self-edges from the unipartite networks. In Table~\ref{networks_info_I}, we show some summary statistics for each of the ten periods ($0607$\_I, $0607$\_II, $0708$\_I, $0708$\_II, $0809$\_I, $0809$\_II, $0910$\_I, $0910$\_II, $1011$\_I, and $1011$\_II). As a basic network statistic, for each bipartite network, we show the degree distributions of bills and legislators in Fig.~\ref{bipartite_degree_distribution}. Observe that the number of legislators who cosponsor bills is distributed much more heterogeneously compared to the number of bills that individual legislators cosponsor. A similar result was reported for legislation cosponsorship networks in the U.S.~\cite{fowler2006b}. From now on, we examine the weighted unipartite networks between legislators [see Fig.~\ref{bipartite_procedure}(b)]. To obtain deeper insights than what we can see from the basic calculations in Fig.~\ref{bipartite_procedure} (and to illustrate a fascinating difference between the Peruvian and U.S. legislatures), we will examine multilayer community structure across the ten legislative sessions. In particular, this will allow us to explore how coherent, densely-connected sets of legislators change from 2006 to 2011.

By partitioning the bills into ten consecutive periods, we obtain a temporal network of cosponsorship relations. The weighted legislator-mode projection networks for different periods all include most legislators (though some legislators are ``invisible'' during a specific period due to their absence in cosponsoring activities). There are 130 legislators\footnote{The 2006--2011 Peruvian Congress had 120 seats, but an additional 10 legislators were part of this Congress because some legislators died or were expelled.} in total, and we examine the temporally changing relations in legislation cosponsorship over the ten periods. We construct a multilayer representation~\cite{mikko,Mucha2010} of the temporal network from the ten consecutive periods. A legislator in a given period is a single node-layer, so each legislator has ten corresponding node-layers. To study community structure, we will partition the multilayer network, which is composed of the node-layers and the edges between them, into disjoint communities.

%%%%%

\subsection{Multilayer community detection}
\label{sec:community}

Reference~\cite{Mucha2010} introduced a method for time-dependent community detection by deriving and maximizing a multilayer generalization of the modularity objective function:
\begin{equation}
	Q = \frac{1}{2\mu} \sum_{ijsr} \left[ \left( A_{ijs} - \gamma_s \frac{k_{i,s}k_{j,s}}{2m_s} \right) \delta(s,r) + 
\delta(i,j) C_{jsr} \right] \delta \left(g_{i,s},g_{j,r}\right) \,,
\label{Q_multilayer}
\end{equation}
where $i$ and $j$ are legislators, $s$ and $r$ (where $s,r \in \{1,\ldots,\mathcal{T}\}$, and we divide time into $\mathcal{T}$ nonoverlapping windows) are layers (i.e., periods, which consist of half-years), $\delta$ is the Kronecker delta, and $g_{i,s}$ is the community to which legislator $i$ is assigned in layer $s$ (so $\delta(g_{i,s}, g_{j,r} ) = 1$ if node $i$ in layer $s$ and node $j$ in layer $r$ belong the same community). The adjacency-tensor element $A_{ijs}$ gives the number of bills that legislators $i$ and $j$ cosponsor in period $s$. Thus, $A_{ijs} = 0$ if $i$ and $j$ do not cosponsor a single bill, and we take $A_{iis} = 0$ for every $i$ to remove self-edges. Additionally, $k_{i,s}$ is the strength of node $i$ (i.e., the sum of the weights of the intralayer edges attached to node $i$) in layer $s$, the quantity $m_s = \sum_{ij} A_{ijs}/2$ is the sum of the weights in layer $s$, and $\gamma_s$ is the resolution parameter in layer $s$. The quantity $C_{jsr} \ne 0$ if node-layer $(j,s)$ is adjacent to node-layer $(j,r)$ (i.e., legislator $j$ is connected to him/herself across different periods), and $C_{jsr} = 0$ otherwise. We assume so-called ``diagonal'' coupling \cite{mikko}, so nonzero interlayer coupling cannot occur between different legislators. We use the factor $2\mu = \sum_{ijs} A_{ijs} + \sum_{jsr} C_{jsr}$ to normalize $Q \in [-1,1]$. See \cite{bazzi2015} for recent theoretical work on multilayer modularity maximization. When a legislator does not appear in some period, all intralayer and interlayer edges attached to his/her associated node-layer have a value of $0$.

We simplify the interlayer connections further by taking $C_{jsr} = \omega$ if two consecutive layers $s$ and $r$ share 
legislator $j$ and $C_{jrs} = 0$ otherwise. Hence, interlayer edges occur only between consecutive layers, the interlayer coupling is ``ordinal'' \cite{mikko}, and it is uniform across legislators. By also taking $\gamma_s = \gamma$ for all layers, we are left with an ``intralayer resolution parameter'' $\gamma$ to go along with the `interlayer resolution parameter' $\omega$ \cite{bassett2013}. As in one of the approaches in Ref.~\cite{Mucha2010}, we use a variant \cite{genlouvain} of the Louvain method~\cite{Blondel2008} to maximize $Q$.

%%%%%%%%

\section{Results}
\label{sec:results}

\begin{figure*}\centering
\begin{tabular}{ll}
(a) & (b) \\
\includegraphics[width=0.45\textwidth]{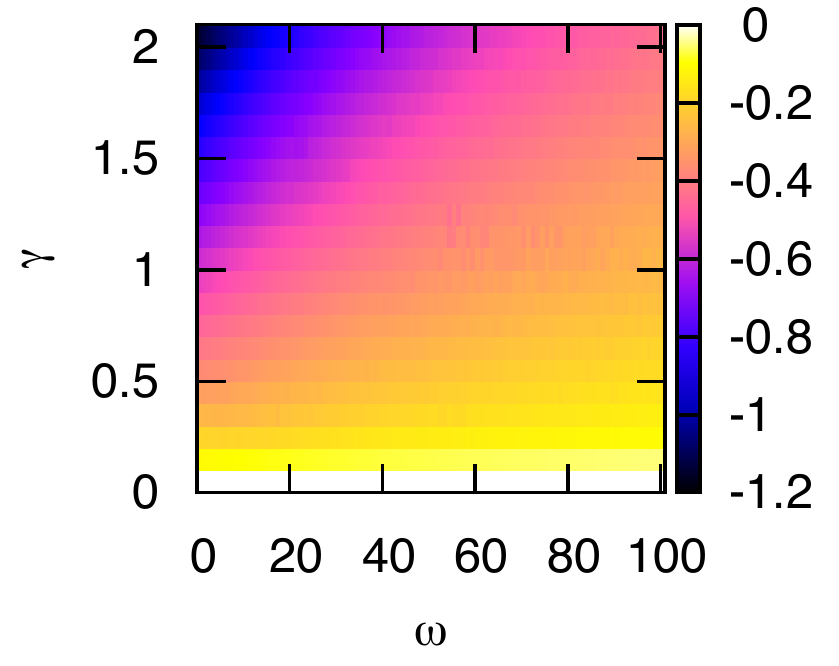} & \includegraphics[width=0.45\textwidth]{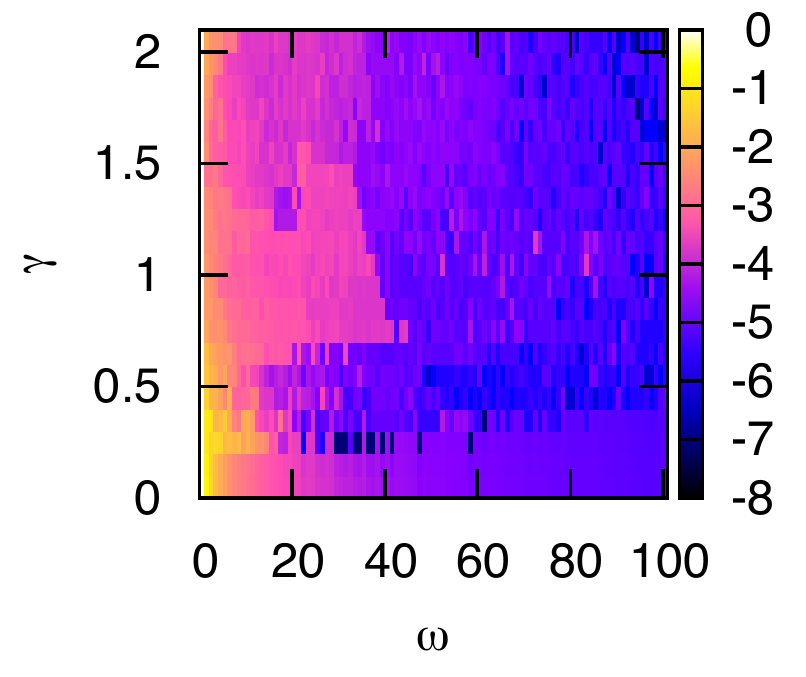} \\
\end{tabular}
\caption{{Summary statistics for time-dependent community detection in the multilayer legislation cosponsorship network in the 2006--2011 Peruvian Congress.} Logarithm of (a) maximum modularity (using a variant \cite{genlouvain} of the Louvain method) and (b) the mean normalized flexibility {$\langle f_i \rangle / (\mathcal{T}-1)$, 
where the flexibility $f_i$ for node $i$ is defined in Eq.~\eqref{eq:flexibility}, $\mathcal{T}$ is the total number of layers, and $\langle \cdots \rangle$ denotes the mean over all nodes.} Flexibility depends on both the interlayer coupling strength (i.e., interlayer resolution parameter) $\omega$ and the intralayer resolution parameter $\gamma$.
}
\label{LPN_multilayer_results}
\end{figure*}

\begin{figure*}\centering
\begin{tabular}{l}
(a) \\
\includegraphics[width=0.95\textwidth]{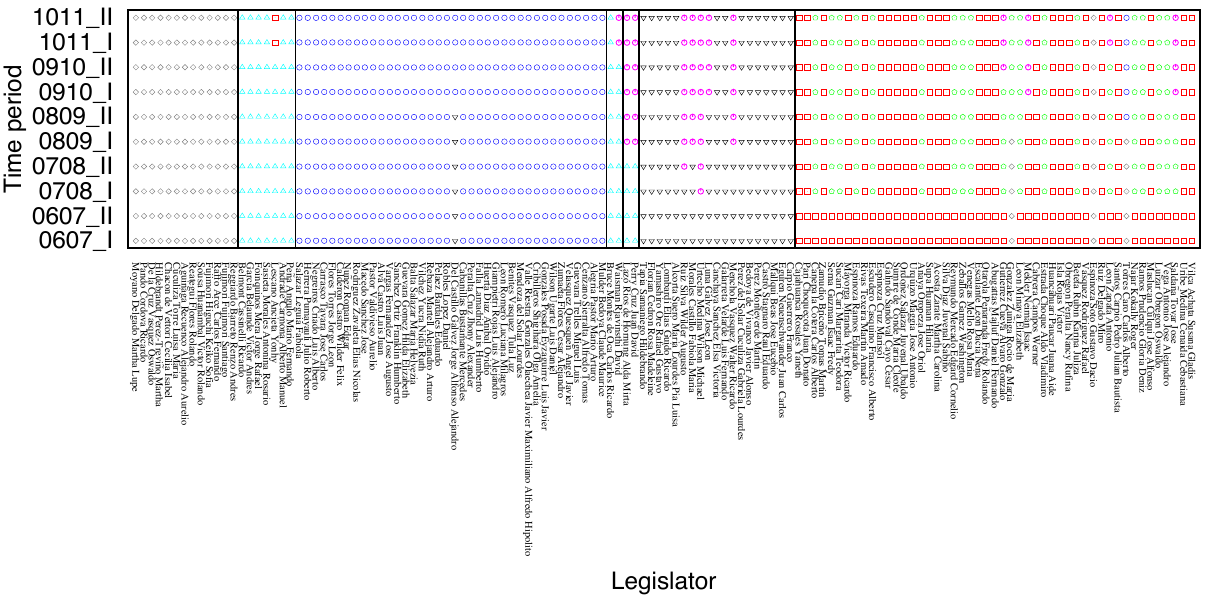} \\
(b) \\
\includegraphics[width=0.95\textwidth]{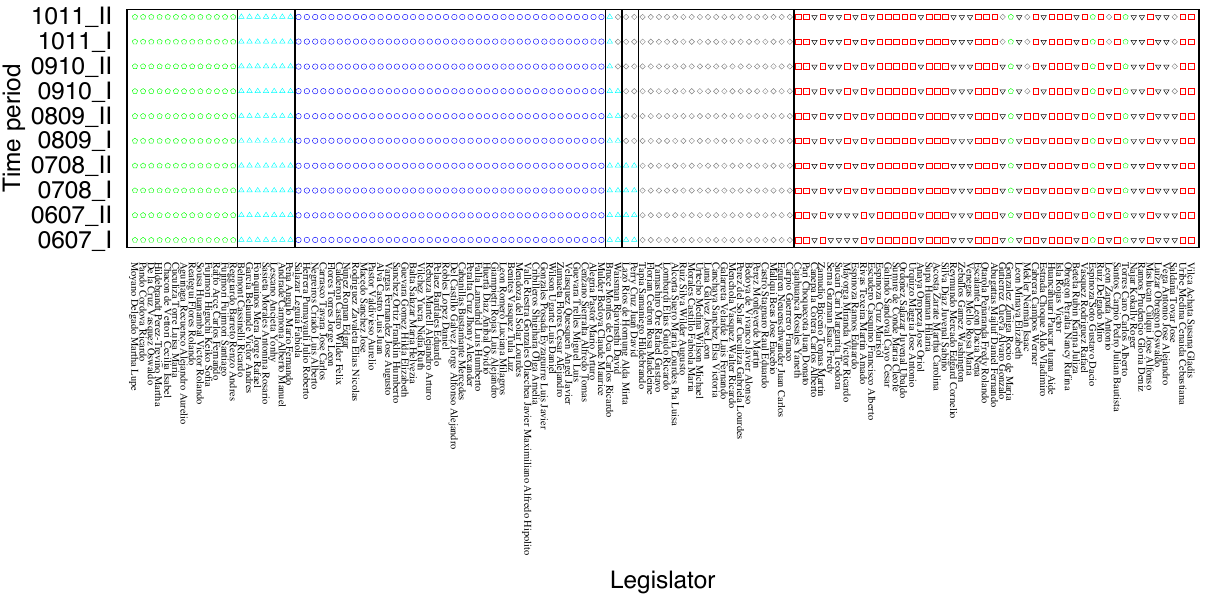} \\
\end{tabular}
\caption{Time-dependent community structure for the multilayer legislation cosponsorship network in the 2006--2011 Peruvian Congress for (a) $(\gamma,\omega) = (1.0,20.0)$ {(with $7$ total communities)} and (b) $(\gamma,\omega) = (1.0,60.0)$ {(with $6$ total communities)}. We show different communities using different colors and symbols, and we sort the legislators by political party (i.e., the parties for which they were candidates), which we sort alphabetically from left to right and separate with vertical lines. The political parties are {Alianza por el Futuro}, {Frente del Centro}, {Partido Asprista Peruano}, {Per{\'u} Posible}, {Restauraci{\'o}n Nacional}, {Unidad Nacional}, and {Uni{\'o}n por el Per{\'u}}.
}
\label{paint_drip_plots_g1}
\end{figure*}

\begin{figure*}\centering
\begin{tabular}{l}
(a) \\
\includegraphics[width=0.95\textwidth]{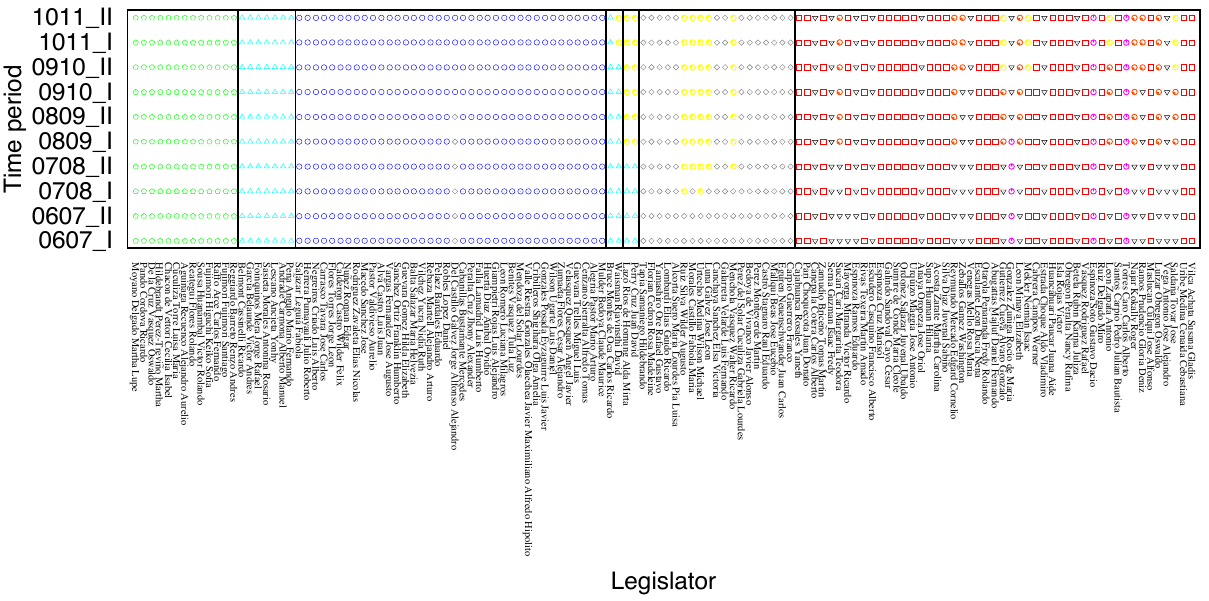} \\
(b) \\
\includegraphics[width=0.95\textwidth]{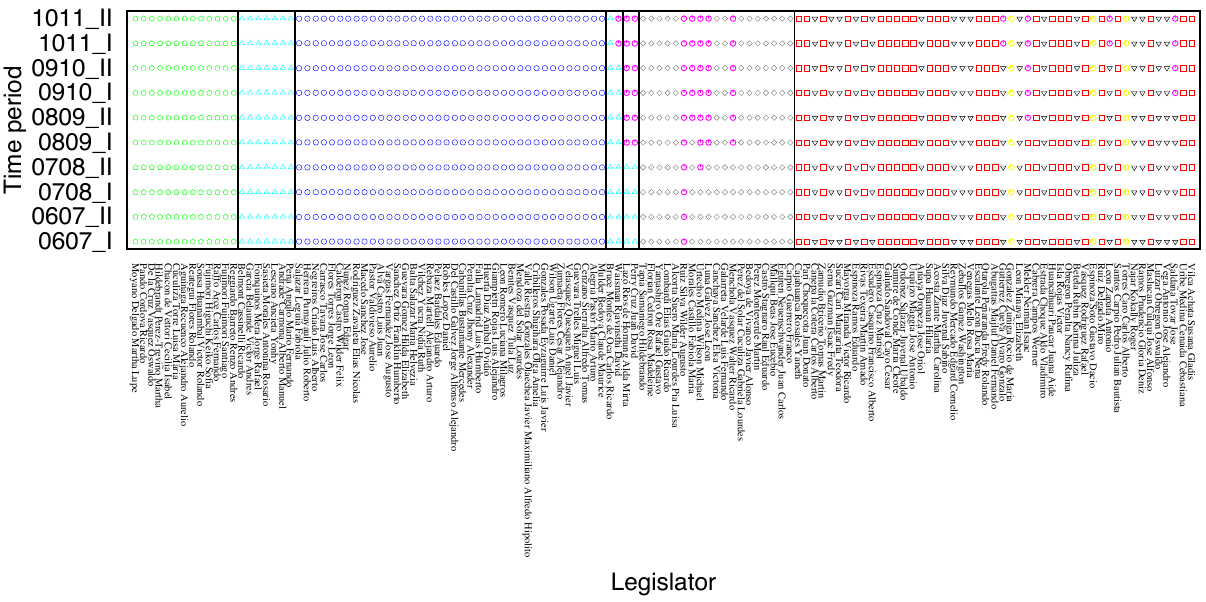} \\
\end{tabular}
\caption{Time-dependent community structure for the multilayer legislation cosponsorship network in the 2006--2011 Peruvian Congress for (a) $(\gamma,\omega) = (1.5,20.0)$ {(with $9$ total communities)} and (b) $(\gamma,\omega) = (1.5,60.0)$ {(with $8$ total communities)}. We show different communities using different colors and symbols, and we sort the legislators by political party, which we sort alphabetically from left to right and separate with vertical lines.The political parties are {Alianza por el Futuro}, {Frente del Centro}, {Partido Asprista Peruano}, {Per{\'u} Posible}, {Restauraci{\'o}n Nacional}, {Unidad Nacional}, and {Uni{\'o}n por el Per{\'u}}.
}
\label{paint_drip_plots_g1_5}
\end{figure*}

\begin{figure*}\centering
\begin{tabular}{l}
(a) \\
\includegraphics[width=0.95\textwidth]{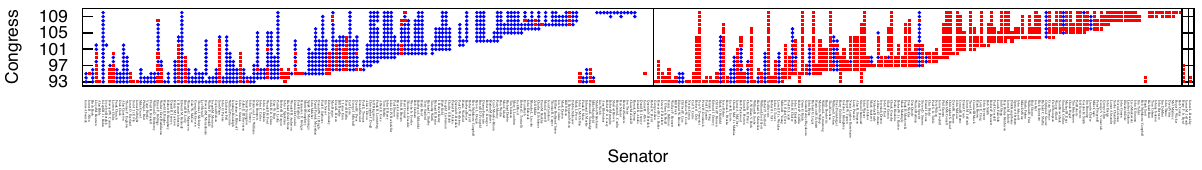} \\
(b) \\
\includegraphics[width=0.95\textwidth]{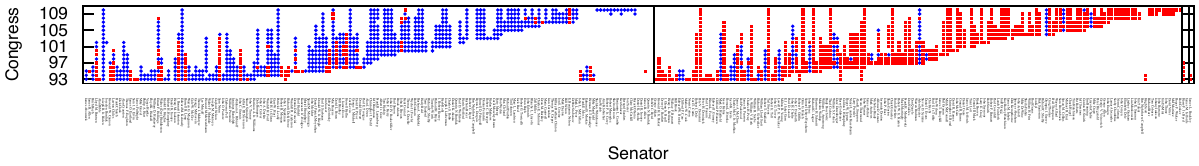} \\
\end{tabular}
\caption{Time-dependent community structure for the legislation cosponsorship network in Congresses 93--110 of the U.S. Senate for (a) $(\gamma,\omega) = (1.0,20.0)$ and (b) $(\gamma,\omega) = (1.0,60.0)$.{ In each case, there are 2 communities in total}. We show different communities using different colors and symbols, and we sort the Senators by political party, which we separate with vertical lines.  From left to right, the parties are the Democratic Party, the Republican Party, independents (Harry F. Byrd Jr. and James M. Jeffords), and the Conservative Party of New York State (James L. Buckley).
}
\label{US_paint_drip_plots_g1}
\end{figure*}

\begin{sidewaysfigure}\centering
\begin{tabular}{ll}
(a) & (b) \\
\includegraphics[width=0.4\textwidth]{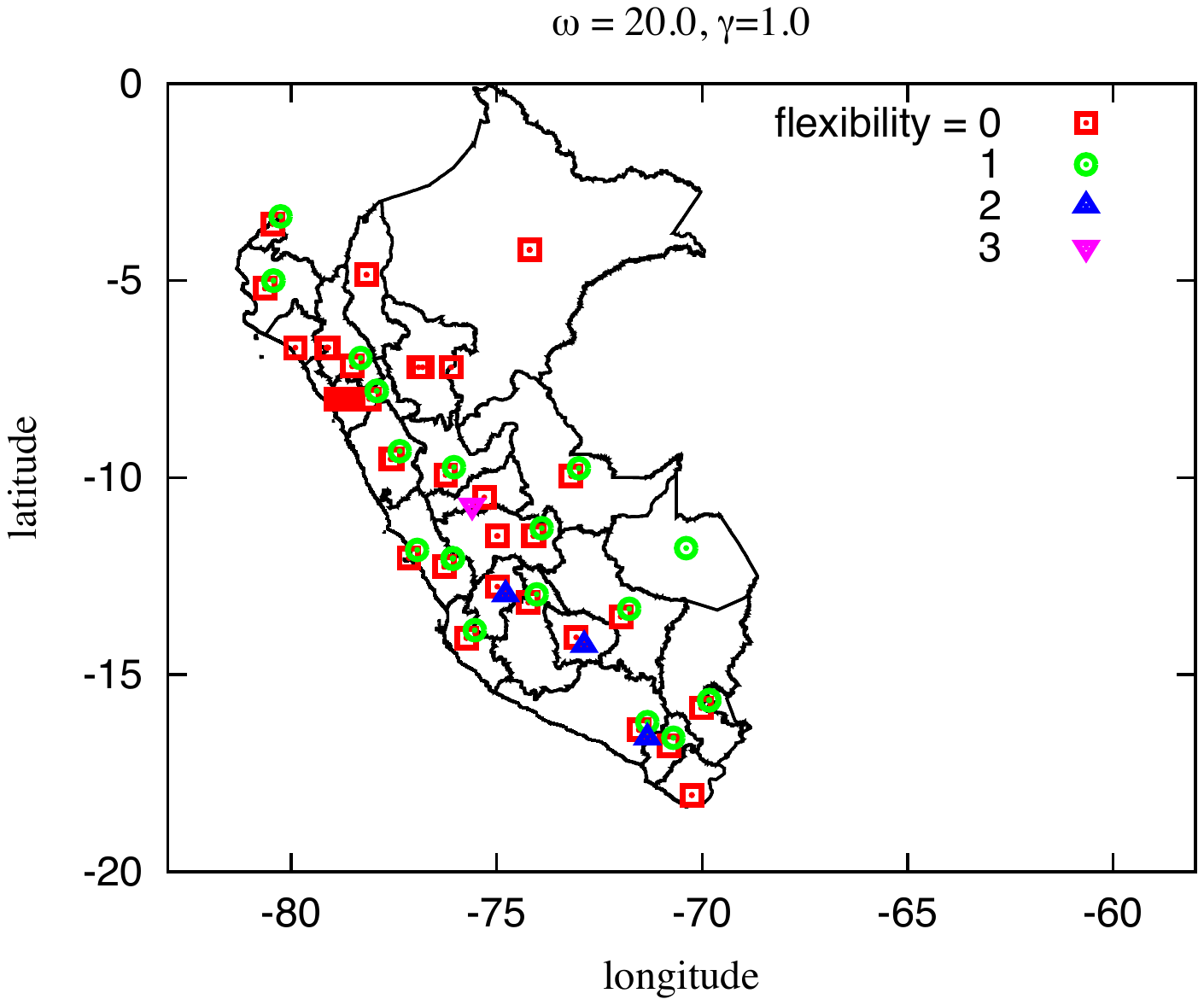} &
\includegraphics[width=0.4\textwidth]{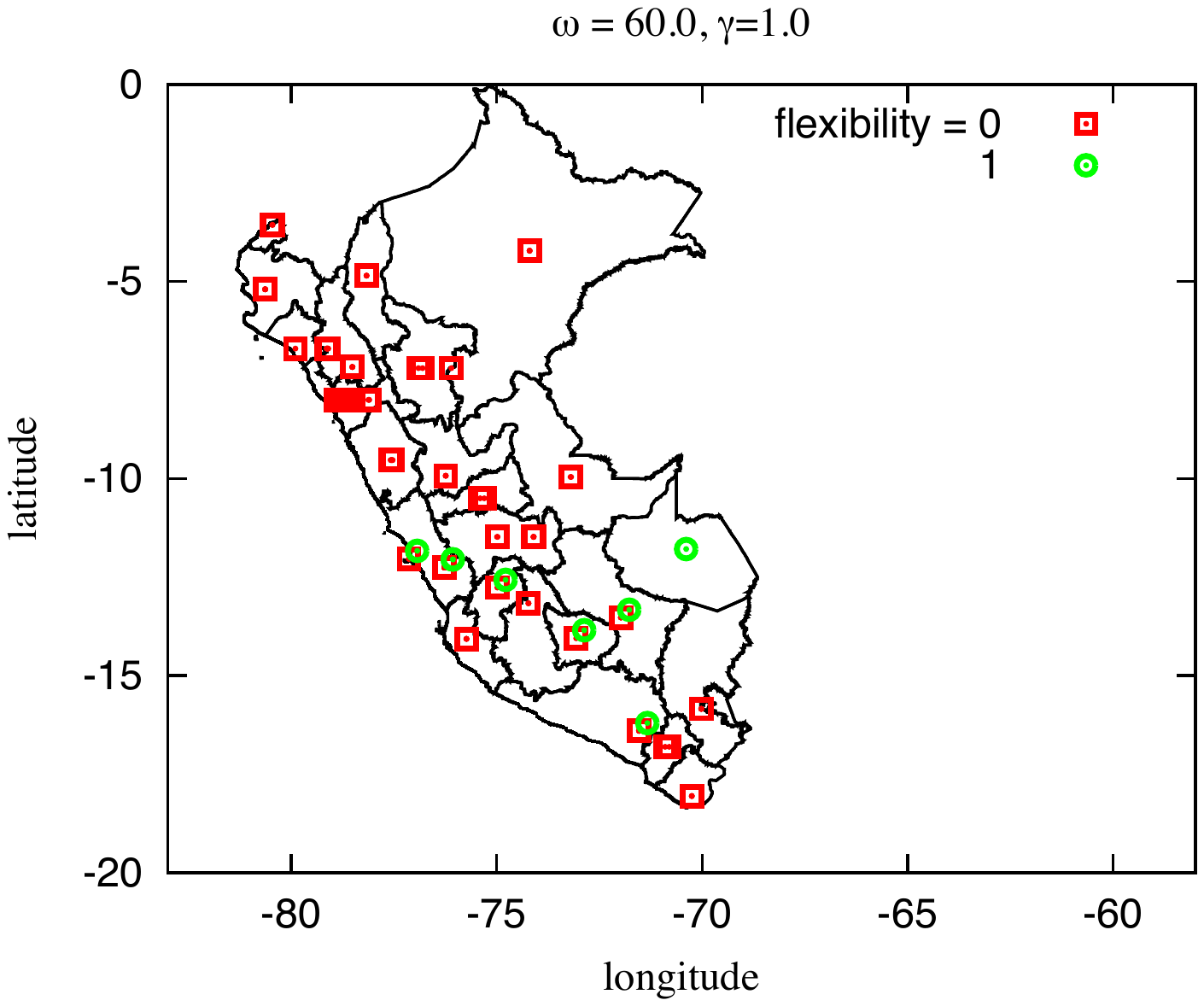} \\
(c) & (d) \\
\includegraphics[width=0.4\textwidth]{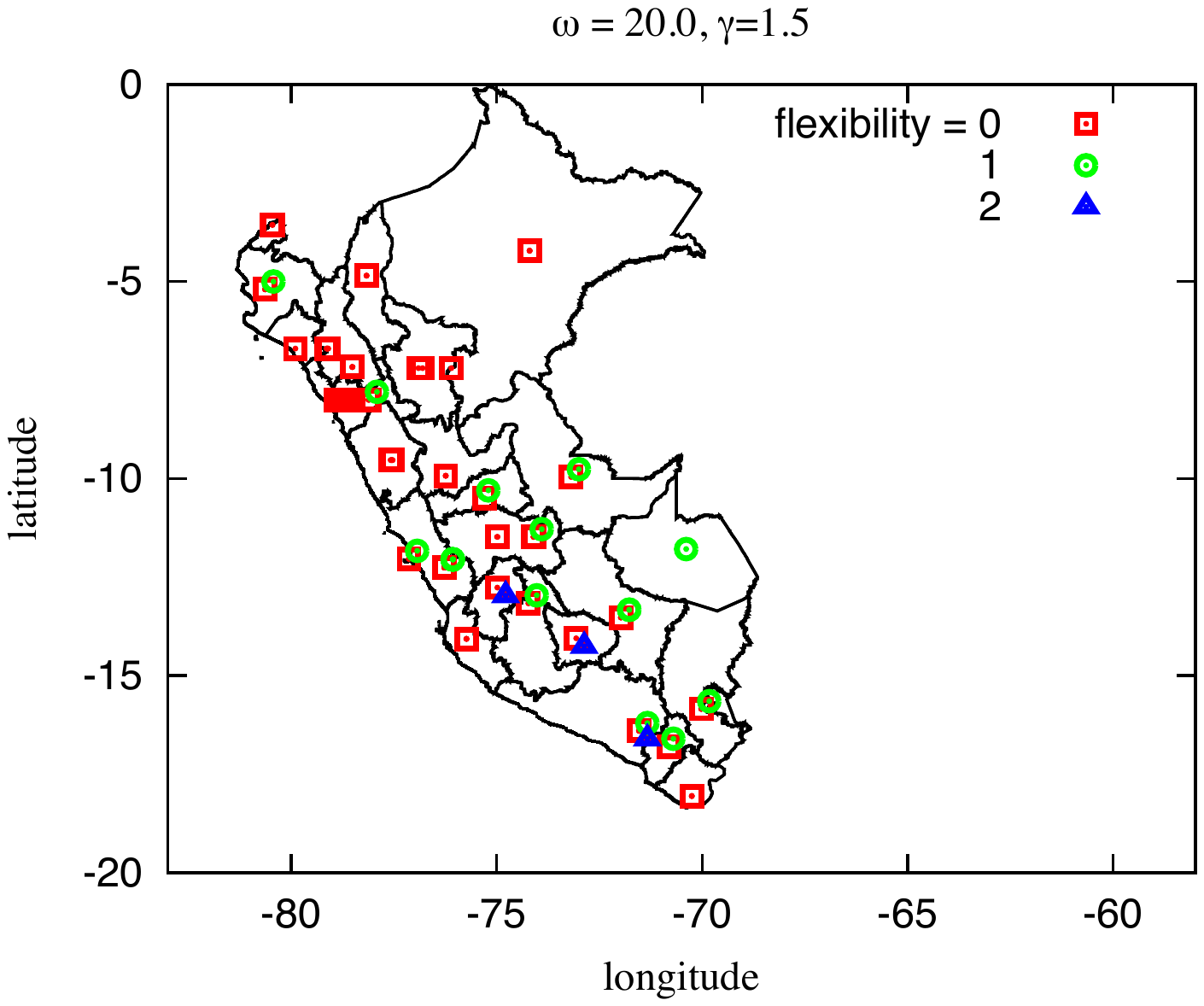} &
\includegraphics[width=0.4\textwidth]{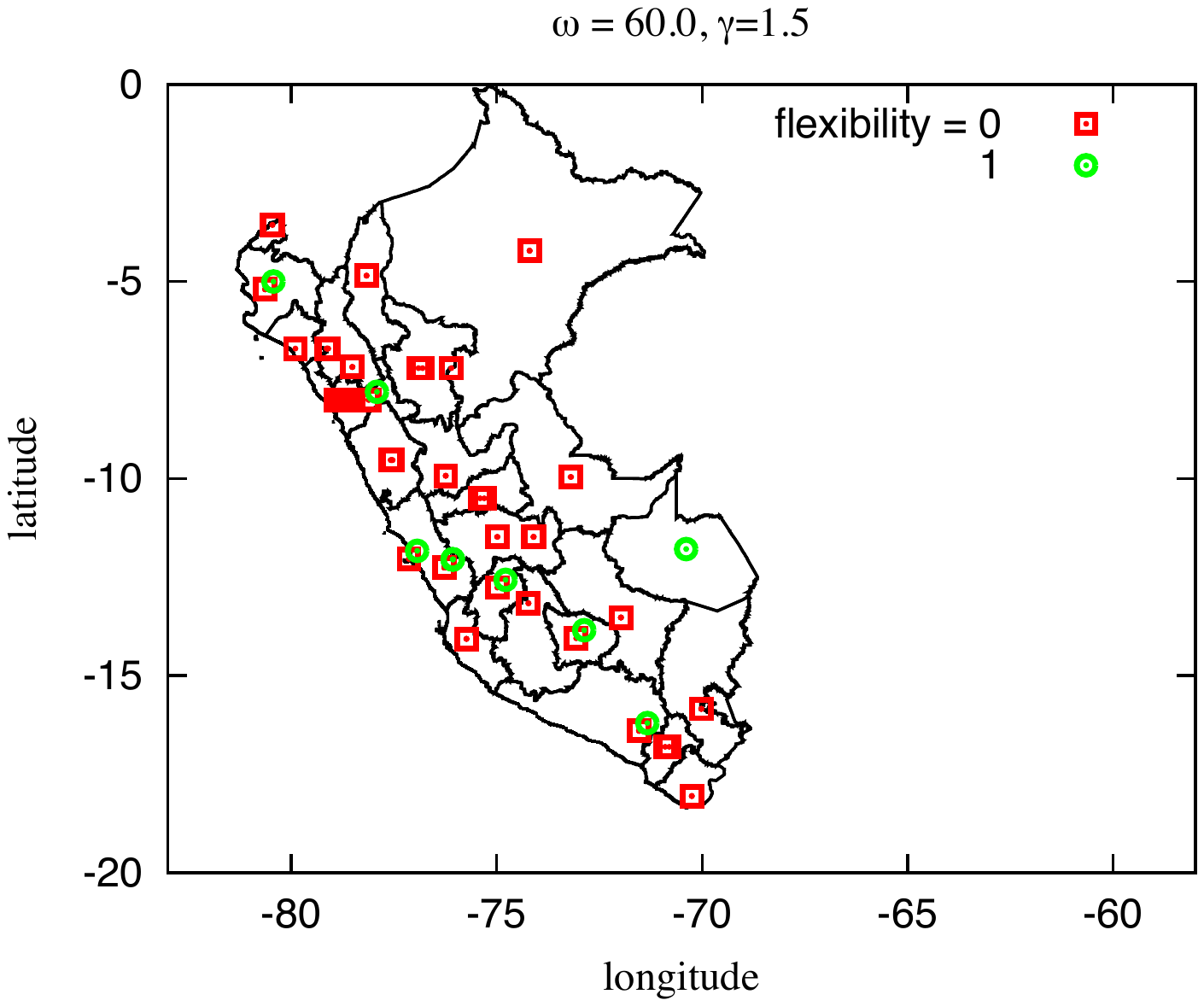} \\
\end{tabular}
\caption{Flexibility values of individual Peruvian legislators on top of their district locations {(where we move the positions slightly differently for different flexibility values to avoid overlap and make the plots more readable)} on the map of Peru~\cite{PeruMap} for (a) $(\gamma,\omega) = (1.0,20.0)$, (b) $(\gamma,\omega) = (1.0,60.0)$, (c) $(\gamma,\omega) = (1.5,20.0)$, and (d) $(\gamma,\omega) = (1.5,60.0)$. The flexibility values are not normalized, so a legislator's flexibility value is the number of times that he/she changed communities during the 2006--2011 Peruvian Congress. 
}
\label{geolocation_flex}
\end{sidewaysfigure}

%\begin{table*}
\begin{sidewaystable}[!p]
\caption{The most flexible Peruvian legislators from the calculations visualized in Figs.~\ref{paint_drip_plots_g1}, \ref{paint_drip_plots_g1_5}, and \ref{geolocation_flex}.} 
%\begin{ruledtabular}
\tiny
\centering
\begin{tabular}{lllcccc}
\hline
 & & & flexibility & & & \\
 name & party & district & with $(\gamma,\omega)=$ & & & \\
 & & & $(1.0,20.0)$ & $(1.0,60.0)$ &
 $(1.5,20.0)$ & $(1.5,60.0)$ \\
\hline
Gloria Deniz Ramos Prudencio & {Uni{\'o}n por el Per{\'u}} & Pasco & $3$ & $0$ & $1$ & $0$ \\
Jos{\'e} Salda{\~n}a Tovar & {Uni{\'o}n por el Per{\'u}} & Huancavelica & $2$ & $1$ & $2$ & $1$ \\
Antonio Le{\'o}n Zapata & {Uni{\'o}n por el Per{\'u}} & Apurimac & $2$ & $1$ & $2$ & $1$ \\
Alvaro Gonzalo Guti{\'e}rrez Cueva & {Uni{\'o}n por el Per{\'u}} & Arequipa & $2$ & $1$ & $2$ & $1$ \\
David Waisman Rjavinsthi & {Per{\'u} Posible} & Lima & $1$ & $1$ & $1$ & $1$ \\
Juan David Perry Cruz & {Restauraci{\'o}n Nacional} & Madre de Dios & $1$ & $1$ & $1$ & $1$ \\
Isaac Mekler & {Uni{\'o}n por el Per{\'u}} & Callao & $1$ & $1$ & $1$ & $1$ \\
Alda Mirta Lazo R{\'i}os de Hornung & {Restauraci{\'o}n Nacional} & Lima & $1$ & $1$ & $1$ & $1$ \\
V{\'i}ctor Ricardo Mayorga Miranda & {Uni{\'o}n por el Per{\'u}} & Cusco & $0$ & $1$ & $1$ & $0$ \\
Walter Ricardo Menchola V{\'a}squez & {Unidad Nacional} & Madre de Dios & $1$ & $0$ & $1$ & $1$ \\
Jos{\'e} Le{\'o}n Luna G{\'a}lvez & {Unidad Nacional} & Lima & $1$ & $0$ & $1$ & $1$ \\
Wilson Michael Urtecho Medina & {Unidad Nacional} & La Libertad & $1$ & $0$ & $1$ & $1$ \\
Fabiola Mar{\'i}a Morales Castillo & {Unidad Nacional} & Piura & $1$ & $0$ & $1$ & $1$ \\
\hline
\end{tabular}
%\end{ruledtabular}
\label{table:geolocation_flex}
%\end{table*}
\end{sidewaystable}

\begin{sidewaysfigure}\centering
\begin{tabular}{ll}
(a) & (b) \\
\includegraphics[width=0.45\textwidth]{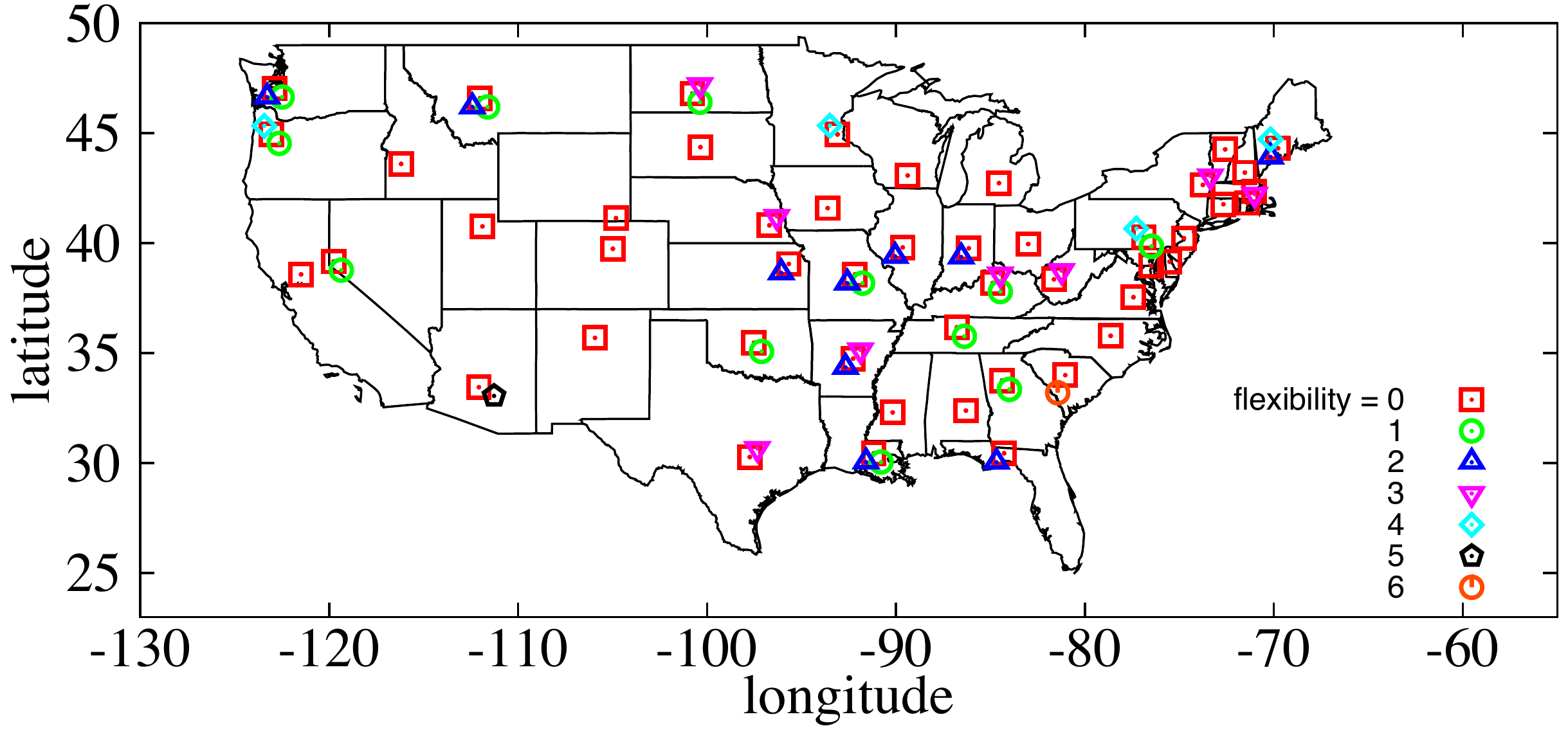} &
\includegraphics[width=0.45\textwidth]{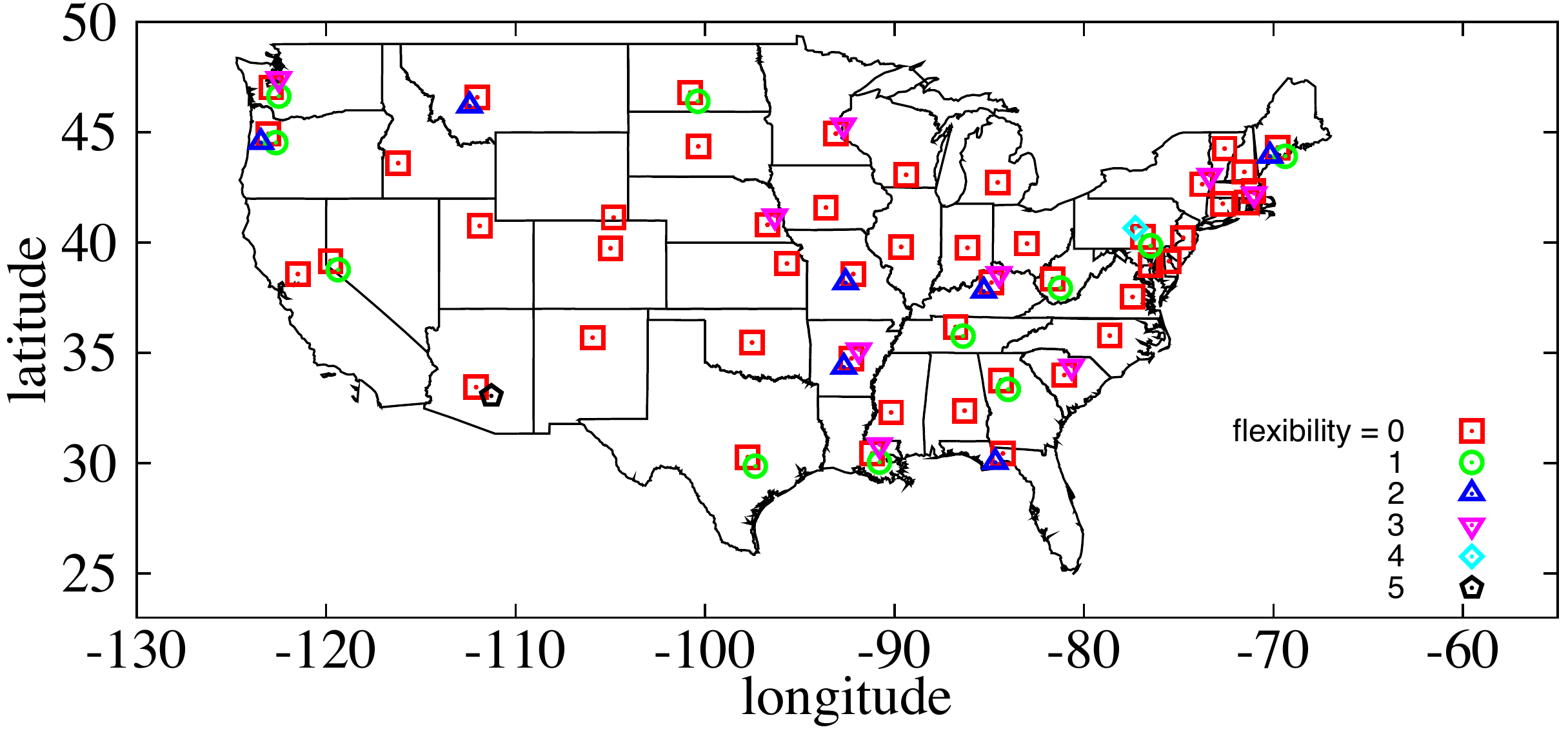} \\
(c) & (d) \\
\includegraphics[width=0.45\textwidth]{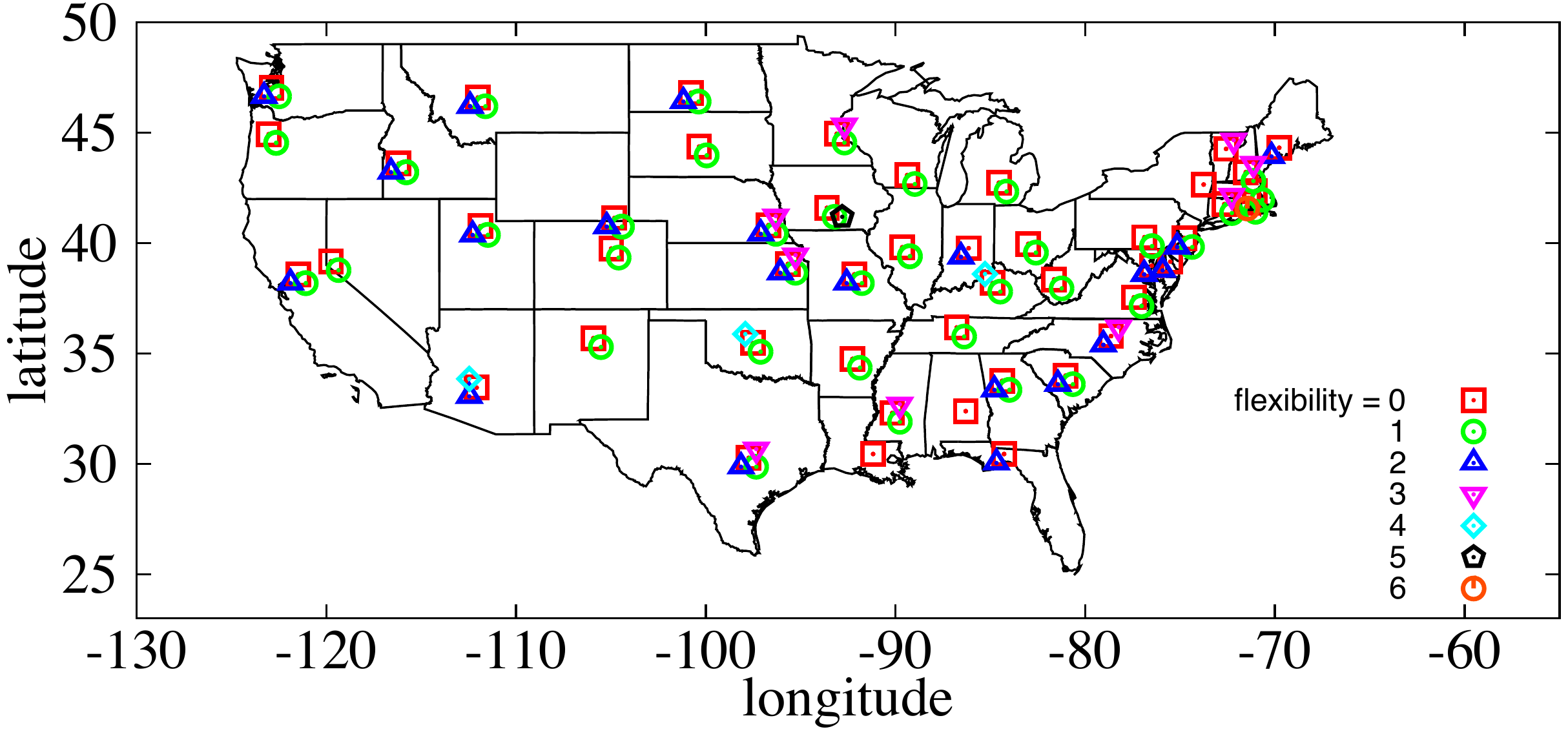} &
\includegraphics[width=0.45\textwidth]{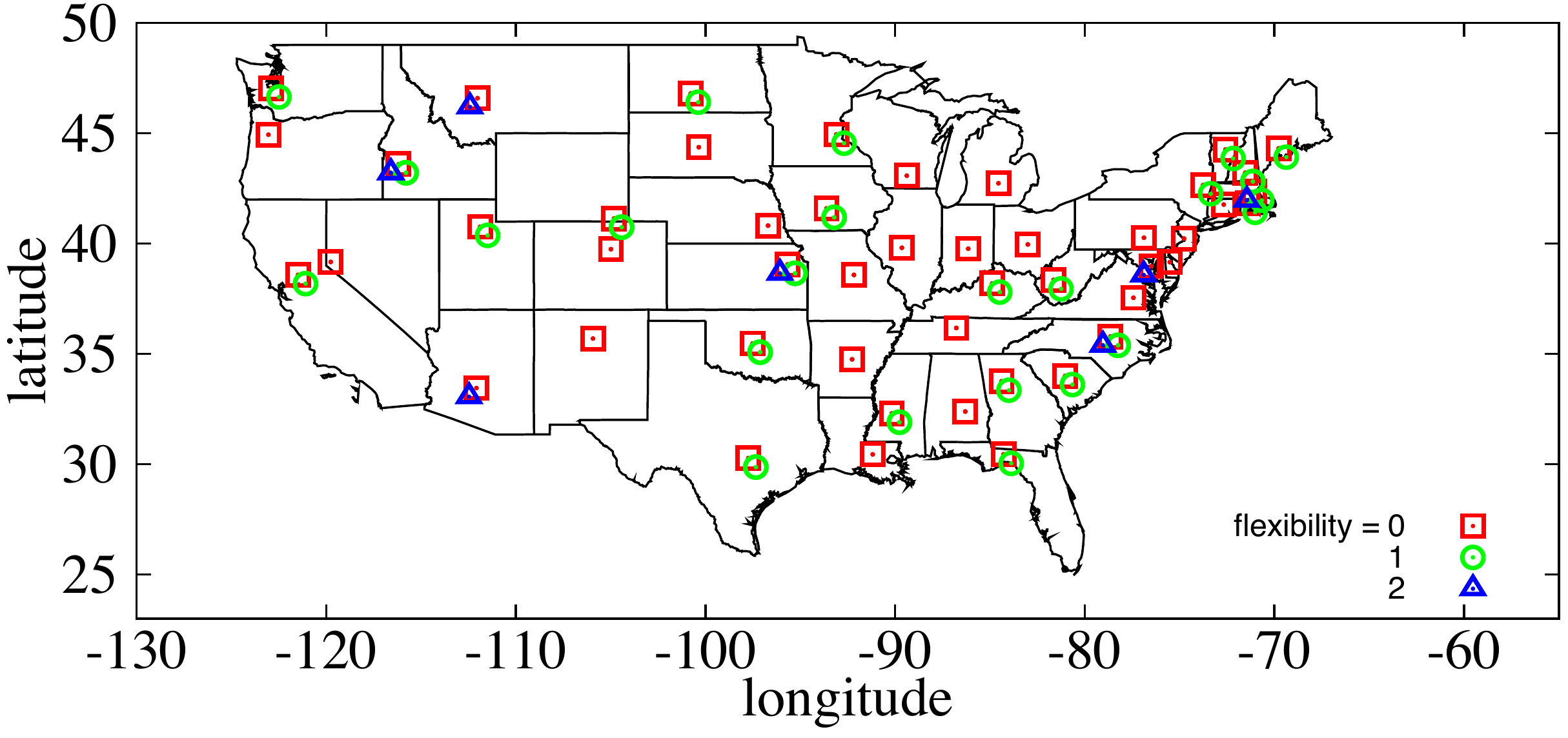} \\
\end{tabular}
\caption{Individual U.S. Senators' flexibility values on top of their states (marked on each state's capital{, where we move the positions slightly differently for different flexibility values to avoid overlap and make the plots more readable)}) on a map of the U.S. mainland~\cite{US_map} for (a) $(\gamma,\omega) = (1.0,20.0)$, (b) $(\gamma,\omega) = (1.0,60.0)$, (c) $(\gamma,\omega) = (1.5,20.0)$, and (d) $(\gamma,\omega) = (1.5,60.0)$. We omit Hawaii and Alaska for ease of viewing. The flexibility values are not normalized, so a Senator's flexibility value is the number of times that he/she changed communities during the 93rd--110th Congresses.
}
\label{geolocation_flex_US}
\end{sidewaysfigure}

%\begin{table*}
\begin{sidewaystable}[!p]
\caption{Notably flexible U.S. Senators from the calculations visualized in Figs.~\ref{US_paint_drip_plots_g1}
and \ref{geolocation_flex_US}.
} 
%\begin{ruledtabular}
\small
\centering
\begin{tabular}{lllcccc}
\hline
 & & & flexibility & & & \\
 name & party & state & with $(\gamma,\omega)=$ & & & \\
 & & & $(1.0,20.0)$ & $(1.0,60.0)$ &
 $(1.5,20.0)$ & $(1.5,60.0)$ \\
\hline
Ernest Hollings & Democrat & South Carolina & $6$ & $3$ & $0$ & $0$ \\
Dennis DeConcini & Democrat & Arizona & $5$ & $5$ & $0$ & $0$ \\
Arlen Specter & Republican & Pennsylvania & $4$ & $4$ & $1$ & $0$ \\
Dave Durenberger & Republican & Minnesota & $4$ & $3$ & $0$ & $0$ \\
William Cohen & Republican & Maine & $4$ & $2$ & $0$ & $0$ \\
Bob Packwood & Republican & Oregon & $4$ & $2$ & $1$ & $0$ \\
Wendell Ford & Democrat & Kentucky & $3$ & $3$ & $1$ & $0$ \\
David Pryor & Democrat & Arkansas & $3$ & $3$ & $0$ & $0$ \\
John Breaux & Democrat & Louisiana & $2$ & $3$ & $0$ & $0$ \\
Alfonse D'Amato & Republican & New York & $3$ & $3$ & $0$ & $0$ \\
John Chafee & Republican & Rhode Island & $3$ & $3$ & $0$ & $0$ \\
James Exon & Democrat & Nebraska & $3$ & $3$ & $1$ & $0$ \\
Slade Gorton & Republican & Washington & $2$ & $3$ & $0$ & $0$ \\
Edward Kennedy & Democrat & Massachusetts & $0$ & $0$ & $6$ & $2$ \\
Tom Harkin & Democrat & Iowa & $0$ & $0$ & $5$ & $1$ \\
John McCain & Republican & Arizona & $0$ & $0$ & $4$ & $2$ \\
Mitch McConnell & Republican & Kentucky & $0$ & $0$ & $4$ & $1$ \\
Don Nickles & Republican & Oklahoma & $0$ & $0$ & $4$ & $1$ \\
Robert Dole & Republican & Kansas & $0$ & $0$ & $3$ & $2$ \\
Jesse Helms & Republican & North Carolina & $0$ & $0$ & $3$ & $2$ \\
Larry Craig & Republican & Idaho & $0$ & $0$ & $2$ & $2$ \\
James McClure & Republican & Idaho & $0$ & $0$ & $1$ & $2$ \\
Barbara Mikulski & Democrat & Maryland & $0$ & $0$ & $2$ & $2$ \\
Conrad Burns & Republican & Montana & $0$ & $0$ & $2$ & $2$ \\
\hline
\end{tabular}
%\end{ruledtabular}
\label{table:geolocation_flex_US}
%\end{table*}
\end{sidewaystable}

%%%%%%

\subsection{Measuring switches in political allegiances}
\label{sec:multilayer}

Motivated by the work in~\cite{bassett2011}, we define the ``flexibility'' $f_i$ of legislator $i$ to be the number of times that he/she changes community membership during the observation time in the multilayer network. That is,
\begin{equation}
	f_i = \sum_{s=1}^{\mathcal{T}-1} \left[ 1 - \delta \left(g_{i,s},g_{i,s+1} \right) \right] \,,
\label{eq:flexibility}
\end{equation}
where we recall that $g_{i,r}$ indicates the community assignment of legislator $i$ in layer $r \in \{1,\ldots,\mathcal{T}\}$. The term ``flexibility'' was used in~\cite{bassett2011} in a study of functional brain networks, and we adopt it for the present paper. 
A complementary concept is network ``persistence,'' which was defined in Ref.~\cite{bazzi2015} as
\begin{equation}
	\sum_{s=1}^{\mathcal{T}-1} \sum_{i=1}^{N} \delta(g_{i,s},g_{i,s+1}) \, ,
\label{eq:persistence}
\end{equation}
where $N$ is the number of entities (e.g., legislators) in a multilayer network.

In Fig.~\ref{LPN_multilayer_results}, we show the maximum modularity and {mean} flexibility $\langle f_i \rangle$ 
 for the temporal multilayer legislator-projected network from multilayer modularity maximization~\cite{Mucha2010} using a variant \cite{genlouvain} of the Louvain method~\cite{Blondel2008}. 
The flexibility results in Fig.~\ref{LPN_multilayer_results}(b) suggest that there is a nontrivial relationship between flexibility values and the parameters $\gamma$ and $\omega$, so it is helpful to explore these results further for several values of $(\gamma,\omega)$. In this paper, we present results for four parameter combinations: $(\gamma,\omega) = (1.0,20.0), (1.0,60.0), (1.5,20.0)$, and $(1.5,60.0)$. {Although the two values of the interlayer coupling strength $\omega$ seem large in comparison to those in previous works}~\cite{Mucha2010,bassett2013}{, we wish to use values that are comparable to intralayer coupling strengths (i.e., the intralayer edge weights), which we take to be the numbers of cosponsored bills.} 
With these computations, we are able to illuminate interesting political structures in the Peruvian Congress and to subsequently contrast them with community structure in the U.S. Senate.
In Figs.~\ref{paint_drip_plots_g1} and \ref{paint_drip_plots_g1_5}, we illustrate time-dependent community structure for several values of $(\gamma,\omega)$. {Larger values of $\gamma$ capture smaller communities (and a larger number of communities), and larger values of $\omega$ capture more temporally coherent communities (as expected)}~\cite{Mucha2010}. 
We then perform similar computations for U.S. Senate cosponsorship networks. We downloaded the U.S. Senate data from ~\cite{US_Senate}. These data start from the 93rd Congress (which covers the dates 3 January 1973--3 January 1975) and go through the 110th Congress (which covers the dates 3 January 2007--3 January 2009). We construct a multilayer U.S. Senate cosponsorship network using the same procedure (see Fig.~\ref{bipartite_procedure}) as for the Peruvian Congress, except that each time window in the former covers an entire 2-year Congress (in contrast to a half-year for Peru). We also use the same values of the resolution parameters $\gamma$ and $\omega$.

{With our calculations, we demonstrate that multilayer modularity maximization can yield insight about time-dependent community structure in real political cosponsorship data, and the adjustable intralayer resolution parameter and interlayer coupling allow one to explore structure at multiple scales. We now compare results from our two data sets, and we interpret our results on the Peruvian Congress in more detail in Section \ref{sec:restructuring}.}

%%%%%

We illustrate time-dependent community structure for the U.S. Senate in Fig.~\ref{US_paint_drip_plots_g1}). Unsurprisingly, we observe a relatively stable bipolar structure (Democrats versus Republicans) in the U.S. Senate, which contrasts sharply with the more intricate temporal community structure in the Peruvian Congress. Although the observation periods --- 5 years versus 36 years for the whole duration, and one half-year versus two years for each layer --- for the two legislatures are different (which is a very important issue to consider in our comparison of the two legislatures), we stress that it is the \emph{shorter} of the two that experiences much more dramatic changes in community structure, further emphasizing the sharp contract in community-structure dynamics in these two countries. For instance, Unidad Nacional (UN), which was composed of four parties, dissolved in 2008~\cite{NationalUnityWikipedia} after eight years of political alliance. This could help explain the restructuring of its year-2007 members that we observe in Figs.~\ref{paint_drip_plots_g1}(a) and \ref{paint_drip_plots_g1_5}(a,b). 

Figures~\ref{paint_drip_plots_g1} and \ref{paint_drip_plots_g1_5} also highlight the loyalty of politicians in the Partido Aprista Peruano (PAP) and the Fujimori Group; none of their members left those groups to join another group. By contrast, the party Uni\'on por el Per\'u (UPP), which started the 0607\_I Congress (in July 2006) with a majority of the seats (45 seats out of 120), lost members to other groups in Congress. By the end of the 1011\_II session in 2011, UPP had only 7 remaining members~\cite{Presentacion}. Figures \ref{paint_drip_plots_g1_5} and (especially) Figure \ref{paint_drip_plots_g1} illustrate the switching of the UPP legislators to other parties.  We discuss such political reorganizations further in Section \ref{sec:restructuring}.

%%%%

\subsection{Dynamical reorganization of the Peruvian Congress}
\label{sec:restructuring}

To discuss time-dependent community structure in the Peruvian Congress in more detail, we need to give some context about the legislators and the Peruvian political system. 
\begin{itemize}
\item{A legislator must obtain support from at least five other legislators to present a bill proposal. Each legislator represents one Region (similar to a U.S. State) in Peru, and each Region can have one or more legislators from one or more parties. Lima, the capital of Peru, has 30\% of all seats \cite{Oficina}. A legislator needs to be part of a political group that includes at least six legislators to be represented on the Congress Board\footnote{{According to the norms of the Congress, available at \cite{norms},}
the Congress Board is in charge of conducting the discussion of proposals, the discussion of voting, and the meetings in which proposals are scheduled. The president of the Congress Board appoints all high-level staff in Congress, and the Congress Board is the body that most influences the Congress's budget.}. 
}
\item{The Peruvian Congress has only one chamber. In 2006, 25 parties competed for 120 seats, and 7 parties won at least one seat~\cite{ELECCIONESGENERALES2006}. The party (PAP) of the President of the executive branch did not obtain a majority (it won 36 seats) in Congress, and the party (UPP) of the runner-up for President won the largest number of seats (45 seats). PAP is a traditional and longstanding party (almost 90 years old) and UPP was composed primarily of new regional leaders and popular figures (from outside the political arena)\footnote{UPP was founded in 1994. After losing against President Alberto Fujimori in 1995, UPP was a weak political party, but it remained eligible to compete in future elections. In the 2006 election, a popular candidate (Ollanta Humala) was invited to be the UPP candidate. See \cite{Historia} for further historical details.}. At the beginning of the 2006--2011 Congress, {it was not clear whether UPP would be able to remain cohesive and gain more support from the minority groups to control the Congress or if the traditional cohesiveness of PAP's legislators would play in PAP's favor and allow them to keep control of Congress.}
}
\end{itemize}

The winning party (PAP) in the executive branch garnered the second-largest number of seats (with 9 fewer than the majority) and
kept control of the Congress. They presided over every Congress Board, {and they never lost their internal cohesion, as none of their members abandoned the party to join another group}.
By contrast, UPP and other parties lost legislators to other political groups or founded new groups. {Among} the 45 legislators that UPP {had} in 2006, only 7 remained in 2011~\cite{GruposParlamentarios}.
Figures~\ref{paint_drip_plots_g1_5} captures reorganizations in the political parties. For instance, we observe opportunistic behavior of legislators who tend to strengthen ties outside their original community. This seems to have occurred not only among legislators whose original party had a small share in Congress but also to members of the majority party (UPP), which included many of the most flexible legislators. Eight legislators were members of at least three different groups; seven of these legislators were from UPP\footnote{These legislators were Gloria Deniz Ramos Prudencio,
Washington Zeballos G\'{o}mez, 
Isaac Mekler Neiman,  
Alvaro Gonzalo Gutierrez Cueva,  
Antonio Le\'on Zapata,  
Jose Salda\~{n}a Tovar,  
and Rosa Mar\'{i}a Venegas Mello.}, and one legislator (Wilson Michael Urtecho Medina) was from the UN. By examining time-dependent community structure, one can also see the emergence of new groups (e.g., Partido Dem\'{o}crata Peruano, which included Carlos Torres Caro, Gustavo Espinoza, and Roc{\'i}o Gonz{\'a}lez from the UPP). One can also observe the cohesiveness in PAP and the Fujimori group (i.e., Alianza por el Futuro).

{Legislators from UPP were not the only ones who were switching to other groups.}
Moreover, the switching behavior of legislators may have been reinforced when UN, {which originally had 17 seats}, ceased to be an alliance in 2008. {The legislators from Solidaridad Nacional (SN),} one of the parties in the UN alliance, {formed a new group 
during the second half of the 2006--2011 Congress (when the 2011 presidential elections were in sight)} \footnote{{The leader of SN, who at that time was the Mayor of Lima (Peru's city capital), aspired to become President in 2011. Meanwhile, other parties in the UN had their own plans for the 2011 Presidential elections.}}.
Officially, however, UPP supported SN in the 2011 election \cite{ENCUENTRA}.

We check for geographical correlations with our observations by comparing individual legislators' district location to their flexibility [see Eq.~\eqref{eq:flexibility}], which indicates how often they change community assignment. In Fig.~\ref{geolocation_flex}, we plot individual legislators' flexibility values on top of their district locations on a map of Peru, and we observe that the central part of the country (including the capital city Lima) tends to have more flexible legislators. We list some notably flexible legislators in Table~\ref{table:geolocation_flex}.

Our discussions above suggest that Peruvian legislators need to be strategic to maximize his/her political opportunities. For example, they can either take a ``loyalist'' or an ``opportunist'' strategy. In the latter strategy, a politician moves to what appears to be a more promising group or party for their political future.
Moreover, one can examine the loyalty level of parties based on how many legislators remain with them over time.
For example, during the 2006--2011 Congress, PAP and the Fujimori group kept all of their members.  Additionally, as we have seen above, having a significant presence in the Peruvian Congress is not sufficient to ensure loyalty: UPP had the majority but lost several legislators to other groups. 
{Such strategic behavior can result from several possible causes:}
\begin{itemize}
    \item{The number of legislators that are needed to found a group is six, which can encourage members of parties with fewer than six seats to switch to other groups (even to ones with a rather different political `ideology' or `identity').}
    \item{There is a current trend of low reelection rates in the Peruvian Congress, so as an election draws near, legislators may need to develop a strategy to distance themselves from parties that become unpopular and associate more closely with groups that become popular.}
    \item{Another current trend is for party leaders to invite popular figures to their Congressional roster. This may affect the cohesiveness of a party.}
\end{itemize}

By 2011, twelve legislators were in a different party than the one to which they were associated during the 2006 elections. Only one of them was reelected in the 2011 election. Nine of the twelve were not from Lima, and none of those nine were reelected. Although party switching is considered to be an effective strategy for short-term gain, the very low reelection rate in the last three Congressional elections (2001, 2006, 2011) may also suggest that it is not an effective long-term strategy. To investigate such a hypothesis, it will be important to compare the Peruvian Congress to Congresses in other nations.

In contrast to party switching in the Peruvian legislature, U.S. Senators have much firmer community memberships, and their flexibility does not seem to include as many geographic characteristics. We show their flexibilities on a map of the U.S. in Fig.~\ref{geolocation_flex_US}, and we again observe a bipolar structure among the U.S. Senators. We list some notably flexible Senators in Table~\ref{table:geolocation_flex_US}. The list depends sensitively on the value of the intralayer resolution parameter $\gamma$ (e.g., it consists mostly of Republicans for some parameter values but not for Democrats for others), in contrast to our observations for the Peruvian Congress. We also recall that using different values of resolution parameters can be helpful for investigating different structural features (e.g., ones with different sizes) in networks \cite{bassett2013,CommunityReview}.

%%%%%

\section{Conclusions and discussion}
\label{sec:conclusion}

We examined time-dependent community structure to explore dynamical restructuring in the Peruvian Congress in the 2006--2011 session by studying networks constructed from (publicly available) legislation cosponsorship data. Our computations give a lens with which to view the frequent switches in the political group affiliations of legislators in Peru,  and our investigation illustrates a dissolution of the majority party at the beginning of 2006 Peruvian Congress. We contrasted these Peruvian politics with the political climate in the U.S., in which there are two highly polarized political parties.

From a more general standpoint, our calculations indicate that even cosponsorship data alone is able to reveal political restructuring in legislative bodies. For countries, such as Peru, in which democratic institutions are still immature,\footnote{Since declaring its independence from Spain in 1821, Peru has had many periods of both democracy and military dictatorships. The current democracy in Peru was reinstated in 1980 after 12 years of military dictatorship, although some scholars have interpreted the government of President Fujimori (1990--2000) as a kind of dictatorship~\cite{Fujimori,692}.} it should be very insightful to use quantitative methods such as network analysis to analyze political restructuring during the maturation process of governmental bodies.

As more data become available amidst the modern data deluge, it will be possible to conduct further
investigations to elucidate both political and social network structures among legislators.
Future research can include focusing on bills rather than politicians, studying change points in the temporal dynamics~\cite{Peel2014}, examining the role of overlapping communities, pursuing case studies in other countries, and more.
It is interesting to examine the relationship between public labels such as party membership 
and quantities (such as flexibility) that one can measure from procedures like time-dependent community detection using (publicly available) legislation cosponsorship data. Ideally, one can use dynamic network analysis to reveal insights that complement those from more traditional techniques (e.g., statistical analyses) in quantitative and qualitative political science.

%%%%%%%

\section*{Acknowledgements}

We thank James Fowler for helpful discussions and for his help in setting up this collaboration. S.\,H.\,L. was supported by the Basic Science Research Program through the National Research Foundation of Korea (NRF) funded by the Ministry of Education (Grant No. 2013R1A1A2011947), and M.\,A.\,P. was supported by the European Commission FET-Proactive project PLEXMATH (Grant No. 317614). S.\,H.\,L. and M.\,A.\,P. also acknowledge support from the Engineering and Physical Sciences Research Council (EPSRC) through grant No. EP/J001759/1.  J.\,M.\,M. acknowledges support  from the University of Washington eScience Institute and data-science grants from the Alfred P. Sloan Foundation, the Gordon and Betty Moore Foundation, and the Washington Research Foundation. He also acknowledges the Center for Social Complexity at George Mason University and the Department of Social Sciences at Pontificia Universidad Cat{\'o}lica del Peru for their support during the initial part of this work. We also thank Peter Mucha for helpful comments.

%%%%%%%

\end{document}